\documentclass[aps,prd,amsfonts,eqsecnum,preprint]{revtex4-1} %
\usepackage{bm}
\usepackage{mathrsfs}
\usepackage{natbib}
\usepackage{graphicx}
\usepackage{amsmath,amssymb,amstext}
\usepackage{subfigure}

\begin{document}
\title{Searching for A Generic Gravitational Wave Background via Bayesian 
Nonparametric Analysis with Pulsar Timing Arrays}
\date{\today}

\author{Xihao Deng}
\email{xud104@psu.edu}
\affiliation{Department of Physics, 104 Davey Laboratory, The Pennsylvania State 
University, University Park, PA  16802-6300, USA}

% flatex input: [abstract.tex]
\begin{abstract}
Gravitational wave background results from the superposition of gravitational 
waves generated from all sources across the Universe. 
Previous efforts on detecting such a background with pulsar timing arrays assume 
it is an isotropic Gaussian background with a power law spectrum. However, when the 
number of sources is limited, the background might be non-Gaussian or the spectrum might
not be a power law. Correspondingly previous 
analysis may not work effectively. 
Here we use a method --- Bayesian 
Nonparametric Analysis --- to try to detect a generic gravitational wave 
background, which directly sets constraints on the feasible shapes of the pulsar 
timing signals induced by a gravitational wave background and allows more 
flexible forms of the background. 
Our Bayesian nonparametric analysis will infer if a gravitational wave 
background is present in the data, and also estimate the parameters that 
characterize the background. This method will be much more effective than the 
conventional one assuming the background spectrum follows a power law in general cases.
While the context of our discussion focuses on pulsar timing arrays, the 
analysis itself is directly applicable to detect and characterize any signals 
that arise from the superposition of a large number of astrophysical events. 
\end{abstract}

% flatex input end: [abstract.tex]

%\usepackage{hyperref}


\maketitle

% flatex input: [intro.tex]
\section{Introduction} \label{sec:intro}

It has been 30 years since Sazhin \cite{sazhin:1978:ofd} showed how 
gravitational waves could be directly
detected by correlating the timing residuals of a collection of pulsars, i.e., a 
pulsar timing array \cite{foster:1990:cap}.  
Such an ``astronomical detector'' is sensitive to gravitational waves of periods 
ranging from the interval between timing observations (weeks to months) to the 
duration of the observational data sets (years), and supermassive black hole 
binaries with masses of $\sim 10^7-10^{10}\mathrm{M_\odot}$ are the primary 
candidates of gravitational wave sources. Gravitational waves generated by a 
large number of such binaries would be superposed to form a background 
\cite{sesana:2008:tsg}. The background has been traditionally assumed as a 
stochastic Gaussian process with a power law spectrum due to Central Limit Theorem 
(e.g.~\cite{hellings:1983:ulo, jenet:2005:dts, jenet:2006:ubo, 
haasteren:2009:omt, haasteren:2011:plo}). %, which states that the sum of a 
%sufficiently large number of independent random variables will be approximately 
%Gaussian distributed. 
This approximation might break down because the gravitational wave contribution 
to the pulsar timing signals may be dominated by the strongest sources, the 
number of which may not be sufficient enough to fulfill the requirement of 
Central Limit Theorem \cite{ravi:2012:das}. %Correspondingly, it is valuable to 
%design a method that is able to detect a generic gravitational wave background 
%without assuming its Gaussianity. 
%
Here we propose a methodology --- Bayesian Nonparametric Analysis 
\cite{ghosh:2003:bn, hjort:2010:bn} --- to analyze the pulsar timing array data 
set that potentially includes contribution from a generic gravitational wave 
background. This method will set strong constraints on the feasible patterns of 
the pulsar timing signals induced by the background we try to search from the 
data. It would investigate if the gravitational wave background is present and also 
estimate the parameters that characterize the background. We will see that when 
the non-Gaussianity of the background becomes non-negligible, our method is 
still efficacious while the conventional method assuming the background is 
Gaussian becomes much less effective.

In Section \ref{sec:char}, we discuss the characteristics of the gravitational 
wave background and its non-Gaussianity. In Section \ref{sec:bayes}, we describe 
a Bayesian nonparametric analysis of the pulsar timing array data. In Section 
\ref{sec:demo}, we illustrate the effectiveness of this analysis by applying it 
to several representative examples and compare the analysis with the 
conventional method assuming the background is Gaussian. Finally, we summarize 
our conclusion in Section \ref{sec:concl}.

% flatex input end: [intro.tex]

%\usepackage{hyperref}

% flatex input: [char.tex]
\section{Characteristics of Gravitational Wave Background} \label{sec:char}
\subsection{Pulsar Timing Residuals Induced by A Gravitational Wave Background}
The evidence of gravitational waves is sought in pulsar timing residuals, which 
are the difference between a collection of pulse ``time of arrival'' (TOA) 
measurements of a pulsar timing array and the predicted pulse arrival times 
based on timing models \cite{hobbs:2010:tip}. A gravitational wave background 
results from the superposition of gravitational waves from a large number of 
sources; correspondingly, pulsar timing response to a gravitational wave 
background would be the sum of timing responses to gravitational waves from 
individual sources.
 
A plane gravitational wave propagating in direction $\hat{k}$ from a single 
source is represented by the transverse-traceless gauge metric perturbation 
\cite{misner:1973:g}
\begin{align}
h_{lm}(t,\vec{x}) &= 
h_{(+)}(t-\hat{k}\cdot\vec{x}) e^{(+)}_{lm} +
h_{(\times)}(t-\hat{k}\cdot\vec{x}) e^{(\times)}_{lm}
\end{align}
where $e^{(A)}_{lm}$ is the polarization tensor. Following \cite{finn:2010:dla}, 
the $j$th pulsar timing response to such a plane gravitational wave can be 
written as

\begin{subequations}\label{eq:inditau}
\begin{equation}
\tau_{j}(t) = \sum_{A=+,\times}\,F^{(A)}_j(\hat{k})\,\left[\tau_{(A)}(t) - 
\tau_{(A)}(
t-L_j(1+\hat{k}\cdot\hat{n}_j))\right]
\end{equation}
where $\tau_{(A)}$ is the integral of $h_{(A)}$ 
\begin{equation}
 \frac{\mathrm{d}\,\tau_{(A)}}{\mathrm{d}\,u} = h_{(A)}(u) 
\end{equation}
and $F^{(A)}$ is the pattern function of the $j$th pulsar,
\begin{equation}
F^{(A)}_j(\hat{k}) = -\frac{\hat{n}^l_j\hat{n}^m_j 
e^{(A)}_{lm}(\hat{k})}{2(1+\hat{k}\cdot\hat{n}_j)}
\end{equation}
where $\hat{n}^l_j$ is the direction from the Earth toward the $j$th pulsar.
\end{subequations}
We can see that the gravitational wave contribution is the sum of two 
functionally identical terms, one time-shifted with respect to the other by an 
amount proportional to the Earth-pulsar distance along the wave propagation 
direction. The first term is referred to as the ``Earth Term'', while the second 
is referred to as the ``Pulsar Term.''

The $j$th Pulsar timing response to a gravitational wave background would be the 
sum of Eq.~\eqref{eq:inditau},
\begin{equation}\label{eq:tauf}
\tau_j(t) = \sum_{a} 
\sum_{A}\,F^{(A)}_j(\widehat{k_{a}})\,\left[\tau^{a}_{(A)}(t) - \tau^{a}_{(A)}(
t-L_j(1+\widehat{k_{a}}\cdot\hat{n}_j))\right]
\end{equation}
where $a$ labels the contribution from the $a$th source. We may express the 
right hand side of Eq.~\eqref{eq:tauf} in terms of its Fourier components:
\begin{align}
\tau_j(t) &= \sum_{a} 
\sum_{A}\,F^{(A)}_j(\widehat{k_{a}})\,\int\mathrm{d}f\,\left[\widetilde{\tau^{a}
_{(A)}}(f)e^{i\,2\pi ft} - \widetilde{\tau^{a}_{(A)}}(
f)e^{i\,2\pi f\left(t-L_j(1+\widehat{k_{a}}\cdot\hat{n}_j)\right)}\right] 
\nonumber \\
&= 
\sum_{A}\int\mathrm{d}f\left[\sum_{a}F^{(A)}_j(\widehat{k_{a}})\widetilde{\tau^{
a}_{(A)}}(f)e^{i\,2\pi ft} - 
\sum_{a}F^{(A)}_j(\widehat{k_{a}})\widetilde{\tau^{a}_{(A)}}(f)e^{-i\,2\pi 
fL_j(1+\widehat{k_{a}}\cdot\hat{n}_j)}e^{i\,2\pi ft}\right]
\end{align}
We notice that compared with the Earth term, the Pulsar term has an extra phase 
$e^{-i\,2\pi fL_j(1+\widehat{k_{a}}\cdot\hat{n}_j)}$. In pulsar timing array 
waveband, the pulsar distance is much longer than the gravitational wavelength, 
i.e., $fL_j \gg 1$. Correspondingly, when summing over the entire sky the Pulsar 
term vanishes and the pulsar timing response to a gravitational wave background, 
i.e., Eq.~\eqref{eq:tauf}, can be simplified as
\begin{equation} \label{eq:tau}
\tau_j(t) = \sum_{a} \sum_{A}\,F^{(A)}_j(\widehat{k_{a}})\,\tau^{a}_{(A)}(t)
\end{equation}

\subsection{Non-Gaussianity of the Gravitational Wave 
Background}\label{sec:nonGaussian}
The assumption of Gaussianity of the gravitational wave background results from 
Central Limit Theorem \cite{feller:1945:tfl}. However, the number of 
gravitational wave sources that can contribute to the pulsar timing array 
waveband is limited, which may not be sufficient to fulfill the requirement of 
the Central Limit Theorem \cite{ravi:2012:das}. Here we use a toy model of 
gravitational wave source population to illustrate how the degree of 
non-Gaussianity increases as the number of gravitational wave sources decreases. 
We refer readers to \cite{ravi:2012:das} for details. 

The degree of non-Gaussianity of a distribution is usually characterized by the 
skewness and kurtosis (e.g.~\cite{joanes:2002:cmo}). The skewness 
$\widehat{S}_X$ and the kurtosis $\widehat{K}_X$ of a random variable $X$ are 
respectively defined as \cite{joanes:2002:cmo}
\begin{subequations}\label{eq:ngauss}
\begin{equation}\label{eq:skewness}
\widehat{S}_X = \frac{\langle \left(X - \langle X \rangle \right)^3 
\rangle}{\left(\langle \left(X - \langle X \rangle \right)^2 
\rangle\right)^{3/2}}
\end{equation}  
\begin{equation} \label{eq:kurtosis}
\widehat{K}_X = \frac{\langle \left(X - \langle X \rangle \right)^4 
\rangle}{\left(\langle \left(X - \langle X \rangle \right)^2 \rangle\right)^{2}} 
- 3
\end{equation}
\end{subequations}
where $\langle\rangle$ denotes ensemble average. If $X$ is Gaussian distributed, 
the skewness $\widehat{S}_X$ and the kurtosis $\widehat{K}_X$ are both zero. 
Negative skewness indicates the left tail of the distribution is longer than the 
right one and positive skewness indicates the right tail is longer than the left 
one \cite{joanes:2002:cmo}; negative kurtosis indicates the distribution has 
shorter tails than Gaussian distribution and positive kurtosis indicates the 
distribution has longer tails than Gaussian distribution \cite{joanes:2002:cmo}. 
If $X$ is the sum of $n$ zero mean identically and independently distributed 
(i.i.d) random variables, i.e., 
\begin{subequations}
\begin{equation}
X = \sum^n_a x_a
\end{equation}
with 
\begin{equation}
\langle x \rangle = 0
\end{equation}
then we can obtain
\begin{align}
\langle X \rangle &= 0 \\ 
\langle (X - \langle X \rangle)^2 \rangle &=  n \langle x^2_a \rangle \\
\langle (X - \langle X \rangle)^3 \rangle &= n \langle x^3_a \rangle  \\
\langle (X - \langle X \rangle)^4 \rangle &= n \langle x^4_a \rangle + 3n(n-1) 
(\langle x^2_a \rangle)^2 
\end{align}
where we have used the fact that the ensemble average of any quantity with a 
linear factor of $x_a$ is zero. Correspondingly, when $n$ is large, we can 
express the skewness and kurtosis as
\begin{align} 
\widehat{S}_X &= \frac{1}{\sqrt{n}}\frac{\langle x^3_a \rangle }{\left( \langle 
x^2_a \rangle\right)^{3/2}} \label{eq:s1} \\
\widehat{K}_X &= \frac{1}{n}\frac{\langle x^4_a \rangle}{\left(\langle x^2_a 
\rangle\right)^{2}} \label{eq:k1}
\end{align}
We can see that as the number of the individual random variable $x_a$ goes to 
infinity, the skewness and kurtosis approach zero and the distribution becomes 
Gaussian. This is what Central Limit Theorem implies. 
\end{subequations}

In the case of gravitational wave background, $X$ is the timing residual 
$\tau_j$ induced by the background in Eq.~\eqref{eq:tau}, and $x_a$ is the 
timing residuals induced by the $a$th single source. For simplicity, we assume 
that the supermassive black hole binaries are the primary sources of 
gravitational waves in pulsar timing array waveband. They are homogeneously and 
isotropically distributed in the sky, and their orbital evolution is driven by 
gravitational wave emission; correspondingly, the number of the binaries per 
frequency per comoving volume is \cite{sesana:2008:tsg}
\begin{equation} \label{eq:density}
\frac{\mathrm{d^2}N}{\mathrm{d}f\,\mathrm{d}V_{\mathrm{c}}} \propto f^{-11/3}
\end{equation}
where $N$ denotes the number of the binaries; $f$ and $V_{\mathrm{c}}$ 
respectively denote gravitational wave frequency and comoving volume of the 
universe. We also assume that the gravitational waves generated by these sources 
are monochromatic waves \cite{ellis:2012:pmf, ellis:2012:osf}. We only consider 
gravitational waves with period $\lesssim 5 \mathrm{yr}$, of which the induced 
peak pulsar timing residuals are above $0.01\,\mathrm{ns}$, since only these 
waves will significantly contribute to the pulsar timing residuals across $5$ 
year observation as the current International pulsar timing array (IPTA) 
\cite{hobbs:2010:tip, demorest:2012:lot, manchester:2012:tpp}. The gravitational 
waves with longer periods will be fitted out by the procedure in the standard 
pulsar timing analysis that removes the linear and quadratic trends induced by 
pulsar spin and spin down. In this range, the number of sources $N$ should be 
$\sim 10^5-10^6$ as expected from theoretical models \cite{sesana:2008:tsg, 
sesana:2011:maw}. 

To compute the skewness and kurtosis of timing residuals induced by a 
gravitational wave background, we can directly follow Eq.~\eqref{eq:s1} and 
Eq.~\eqref{eq:k1}. The timing residual $\tau^a_j$ of the $j$th pulsar induced by 
the sinusoidal gravitational waves generated from the $a$th individual source 
would be 
\begin{subequations}
\begin{align}
\tau^a_j(t) &= F^{(+)}_j(\widehat{k_{a}})\,B^{a}_{(+)}\cos\left(2\pi f^a t + 
\phi^a \right) + F^{(\times)}_j(\widehat{k_{a}})\,B^{a}_{(\times)}\sin\left(2\pi 
f^a t + \phi^a \right)    \nonumber \\
&= B^a_j \cos\left(2\pi f^a t + \phi^a + \psi^a_j \right)
\end{align} 
where $B^a_{(A)}$ denotes the amplitudes of the timing residuals induced by 
$(A)$ polarization component; $f^a$ and $\phi^a$ are respectively the frequency 
and initial phase of the gravitational waves; $B^a_j$ and $\psi^a_j$ are 
respectively 
\begin{align}
B^a_j &= \sqrt{[F^{(+)}_j\,B^{a}_{(+)}]^2 + 
[F^{(\times)}_j\,B^{a}_{(\times)}]^2} \\
\psi^a_j &= 
\arctan\left[\frac{F^{(+)}_j\,B^{a}_{(+)}}{F^{(\times)}_j\,B^{a}_{(\times)}}
\right]
\end{align}
\end{subequations}
Assuming that the initial phase $\phi^a$ is uniformly distributed between $0$ 
and $2\pi$, we can obtain
\begin{subequations}
\begin{align}
\langle \tau^a_j \rangle &= \langle B^a_j\rangle \, \langle \cos\left(2\pi f^a t 
+ \phi^a + \psi^a_j \right) \rangle = 0 \\
\langle \left(\tau^a_j\right)^2 \rangle &= \langle \left(B^a_j\right)^2 \rangle 
\, \langle \cos^2\left(2\pi f^a t + \phi^a + \psi^a_j \right) \rangle = \pi\, 
\langle \left(B^a_j\right)^2 \rangle \\
\langle \left(\tau^a_j\right)^3 \rangle &= \langle \left(B^a_j\right)^3 \rangle 
\, \langle \cos^3\left(2\pi f^a t + \phi^a + \psi^a_j \right) \rangle = 0 \\
\langle \left(\tau^a_j\right)^4 \rangle &= \langle \left(B^a_j\right)^4 \rangle 
\, \langle \cos^4\left(2\pi f^a t + \phi^a + \psi^a_j \right) \rangle = 
\frac{3\pi}{4}\, \langle \left(B^a_j\right)^4 \rangle
\end{align}
and following Eq.~\eqref{eq:s1} and Eq.~\eqref{eq:k1}, the skewness and kurtosis 
of of timing residuals induced by a gravitational wave background would be
\begin{align}
\widehat{S}_\tau &= 0 \\
\widehat{K}_\tau &= \frac{1}{n}\frac{3}{4\pi}\frac{\langle \left(B^a_j\right)^4 
\rangle}{\left[\langle (B^a_j)^2 \rangle\right]^{2}} 
\end{align}
\end{subequations}
The amplitude $B^a_j$ depends on the pattern function, gravitational wave 
amplitudes and frequency. We can compute $\langle \left(B^a_j\right)^2 \rangle$ 
and $\langle \left(B^a_j\right)^4 \rangle$ by sampling an ensemble of the 
amplitude $B^a_j$ from the distribution Eq.~\eqref{eq:density} and numerically 
computing the ensemble average. Table~\ref{tab:ngauss} presents the skewness and 
kurtosis of timing residuals of PSR J1713+0747 induced by gravitational wave 
backgrounds respectively generated from $10^6$, $5\times 10^5$, $2\times 10^5$ 
and $10^5$ sources. As we stated above, all of these sources are generated in 
pulsar timing array waveband, i.e., their gravitational wave periods range from 
$3$ months to $5$ years. Their frequency distribution follows 
Eq.~\eqref{eq:density}. For all these three gravitational wave backgrounds, 
about $0.1\%$ of the sources are responsible for $95\%$ of the residual power.

\begin{table}[ht]
\centering
\begin{tabular}{c  c  c}
\hline \hline
No. of Sources \hspace{1cm} & skewness \hspace{1cm} & kurtosis \\
\hline
$10^6$ \hspace{1cm} & 0 \hspace{1cm} & 0.01 \\
$5\times 10^5$ \hspace{1cm} & 0 \hspace{1cm} & 0.02 \\
$2\times 10^5$ \hspace{1cm} & 0 \hspace{1cm} & 0.05 \\
$10^5$ \hspace{1cm} & 0 \hspace{1cm} & 0.1 \\
\hline
\end{tabular}
\caption[skewness and kurtosis of timing residuals induced by a gravitational 
wave backgrounds]{skewness and kurtosis of timing residuals induced by 
gravitational wave backgrounds respectively generated from $10^6$, $5\times 
10^5$, $2\times 10^5$ and $10^5$ sources.}
\label{tab:ngauss}
\end{table}
We can see that as the number of sources decreases, the degree of 
non-Gaussianity increases as implied by Central Limit Theorem. Correspondingly,
the power law spectrum of the gravitational wave background as generally assumed 
would be effectively modified as it is derived based on the assumption that the 
number of sources are approximately infinite (e.g.~\cite{flanagan:1993:sot, jenet:2005:dts, 
haasteren:2009:omt}).

% flatex input end: [char.tex]

%\usepackage{hyperref}

% flatex input: [bayes.tex]
\section{Bayesian Nonparametric Analysis} \label{sec:bayes}
To seek for the evidence of a gravitational wave background in the data set, we 
need to model its contribution to pulsar timing residuals. In principle, we can 
parameterize the pulsar timing response to a gravitational wave background by 
Eq.~\eqref{eq:tau}, i.e., superposition of gravitational waves from a large 
number of sources. In this way, the gravitational wave background is treated as 
a deterministic process rather than a stochastic one. However, such a parametric 
model will be computationally impossible due to the large number of parameters, 
because the gravitational wave signal induced by one single source requires 
about 10 to 20 parameters to characterize \cite{ellis:2012:pmf, ellis:2012:osf}, 
and correspondingly, to characterize a gravitational wave background generated 
by $\sim 10^6$ sources, the number of parameters will be $\sim 10^7$. To avoid 
both over-parameterization and strong assumption of Gaussianity, we introduce a 
different method --- Bayesian nonparametric analysis \cite{ghosh:2003:bn, 
rasmussen:2006:gpf, hjort:2010:bn}. This method also treats the gravitational 
wave background as a {\it deterministic} process since it originates from the 
superposition of gravitational waves from a finite number of sources, each of 
which is a deterministic process. However, we do not try to write down the exact 
deterministic function form. Instead, we assign a prior distribution on the 
function form of the signal, which will characterize the expected shape of the 
signal pattern, such as its smoothness, variation, trend, etc. These 
characteristics are used to represent the function form of the signal. 
Correspondingly, we are able to detect the signal whose deterministic function 
form has the same characteristics as what our prior distribution characterizes. 
For discussion of application of Bayesian nonparametrics in gravitational wave 
context, see \cite{deng:2014:sfg, lentati:2014:beo, lee:2014:mao} for details. 

\subsection{Framework of Bayesian Nonparametric Analysis} \label{sec:frame}
To infer the pulsar timing residuals $\bm{\tau}$ induced by the gravitational 
wave background  from a pulsar timing array data set $\bm{\mathrm{d}}$, we need 
to compute the posterior probability density $p(\bm{\tau}|\bm{\mathrm{d}})$, 
i.e., probability density of $\bm{\tau}$ given the data set $\bm{\mathrm{d}}$ 
\cite{gelman:2004:bda}. In this paper, we neglect other pulsar timing effects 
such as pulsar spin and spin down that would contribute to the data set 
$\bm{\mathrm{d}}$, and $\bm{\tau}$ only denotes the contribution from the 
gravitational wave background. It is straightforward to include other timing 
effects in our analysis by just adding a timing model to $\bm{\tau}$ like the 
way applied in \cite{haasteren:2009:omt}.

Exploiting Bayes' Theorem, the posterior probability density $p$ can be 
expressed in terms of the likelihood function $\Lambda$
, an a priori probability density $q$ that expresses the expectations of 
$\bm{\tau}$, and a normalization constant $Z$,
\begin{equation} \label{eq:postertau}
p(\bm{\tau}|\bm{\mathrm{d}}) = \frac{\Lambda(\bm{\mathrm{d}}|\bm{\tau})\,
q(\bm{\tau})}{Z(\bm{\mathrm{d}})}
\end{equation}
where $\Lambda(\bm{\mathrm{d}}|\bm{\tau})$ is the likelihood function, which 
describes the probability of the data set $\bm{\mathrm{d}}$ given the signal 
characterized by $\bm{\tau}$; $q(\bm{\tau})$ is the a priori probability density 
of $\bm{\tau}$ that expresses our expectation before we obtain the data set; 
$Z(\bm{\mathrm{d}})$ is the normalization constant. 

As shown in \cite{finn:2010:dla, deng:2014:sfg}, the likelihood function is a 
multivariate Gaussian distribution of the data set $\bm$, as the pulsar timing 
noise is well modelled as zero mean Gaussian distributions and they are 
uncorrelated among different pulsars, i.e.,
\begin{align} \label{eq:likelihood}
\Lambda(\bm{\mathrm{d}}|\bm{\tau}) &= \prod^{N_p}_{j=1} 
\frac{\exp\left[-\frac{1}{2}(\mathrm{d}_j-\tau_j)^{T}\mathrm{C}^{-1}_j(\mathrm{d
}_j-\tau_j)\right]}{\sqrt{(2\pi)^{\mathrm{dim}\,\mathrm{d}_j}\det||\mathrm{C}
_j||}} \nonumber \\
&= N(\bm{\mathrm{d}} - \bm{\tau}|\bm{\mathrm{C}})
\end{align}
where $N_p$ is the number of pulsars in the pulsar timing array, and 
$\mathrm{C}_j$ is the noise covariance matrix of the $j$th pulsar.

The a priori probability density $q(\bm{\tau})$ describes our expectations of 
the pulsar timing residuals induced by the gravitational wave background before 
analyzing the data set $\bm{\mathrm{d}}$, which will set a strong constraint on 
all possible forms of $q(\bm{\tau})$ we try to explore \cite{Sudderth:2006:gmf}.
Under the assumption that we do not have information on the values of 
$\bm{\tau}$ at initial observation time, i.e., the prior of $\bm{\tau}$ holds 
time invariance symmetry, it can be proved that the prior distribution of 
$\bm{\tau}$ is a stationary Gaussian process with zero mean 
\cite{summerscales:2008:mef, deng:2014:sfg, Bretthorst:1988:bsa}
\begin{subequations}
\begin{align} \label{eq:prior0}
q(\bm{\tau}) &= \frac{\exp\left[-\frac{1}{2}\sum\limits_{j,k}\int 
\mathrm{d}t\,\mathrm{d}t'\,\tau_j(t)\,\mathrm{K}^{-1}_{jk}(t,t')\,
\tau_k(t')\right]}{\sqrt{(2\pi)^{\mathrm{dim}\mathrm{K}}\det||\mathrm{K}||}} 
\end{align}
where $j,\, k$ are pulsar indices; $\mathrm{K}$ is a covariance function or 
kernel of the stationary Gaussian process only depending on the time difference 
\cite{rasmussen:2006:gpf}, i.e., 
\begin{align}\label{eq:kernel0}
&\mathrm{K}_{jk}(t,t') = \mathrm{K}_{jk}(\Delta t)
\end{align}
where $\Delta t$ denotes
\begin{equation} 
 \Delta t = |t - t'|
\end{equation}

\end{subequations}
The kernel will characterize the smoothness, trend and variation of the timing 
residual induced by the gravitational wave background and correspondingly set 
strong constraints on the possible shapes of it. In general, there would be some 
nuisance parameters in the kernel, which are referred to as {\it 
hyperparameters} \cite{gelman:2004:bda}. To do a full Bayesian analysis, we also 
need to assign the prior for the hyperparameters, i.e., hyperpriors 
\cite{gelman:2004:bda}. The choice of kernel and hyperprior will be discussed in 
the next subsection.

\subsection{The choice of Gaussian Process Prior}
\subsubsection{Prior of $\boldsymbol{\tau}$} \label{sec:tauprior}
As discussed above, the prior probability of the timing residuals induced by the 
gravitational wave background will constrain their feasible forms we try to 
explore, so the prior we choose should express the characteristics implied in 
Eq.~\eqref{eq:tau}. 

Following Eq.~\eqref{eq:tau}, the timing residual induced by the superposition 
of gravitational waves from direction $\hat{k}$ can be written as
\begin{equation} \label{eq:tauk}
\tau_{\hat{k}}(t) =  
\sum_{A}\,F^{(A)}(\hat{k})\,\sum_{a\,(\hat{k})}\tau^{a}_{(A)}(t)
\end{equation}
where $a\,(\hat{k})$ labels the sources located in $\hat{k}$ direction. 
To choose an appropriate Gaussian process prior of $\tau_{\hat{k}}$, we need to 
first find a Gaussian process prior of $\tau^{a}_{(A)}$ of a single source. 

In pulsar timing array waveband, the single gravitational wave source could be a 
circular binary, an eccentric binary or a burst \cite{hobbs:2010:tip}. The 
dynamics of those gravitational wave sources are expected to be smooth 
\cite{hobbs:2010:tip} and correspondingly, $\tau^{a}_{(A)}$ is expected to be a 
smooth function of time \cite{deng:2014:sfg}. As a consequence, the mean square 
of $\tau^{a}_{(A)}$ under its prior has to be infinitely differentiable, which 
requires the kernel of its stationary Gaussian process prior infinitely 
differentiable at $\Delta t = 0$ \cite{adler:1981:tgo}. Only few kernels we know 
satisfy this requirement \cite{stein:1999:ios} and the one with the least number 
of hyperparameters is \cite{rasmussen:2006:gpf, deng:2014:sfg}
\begin{subequations} 
\begin{equation} \label{eq:kernelS}
 \mathrm{K}_{a(A)}(\Delta t) = \sigma^{2}_{a(A)}\exp(-\frac{\Delta 
t^2}{2\,\lambda_{a}^2})
\end{equation}
and the corresponding Gaussian process prior of $\tau^{a}_{(A)}$ would be 
\begin{equation}
q(\tau^{a}_{(A)}) = \frac{\exp\left[-\frac{1}{2}\int 
\mathrm{d}t\,\mathrm{d}t'\,\tau^{a}_{(A)}(t)\,\mathrm{K}^{-1}_{a(A)}(\Delta 
t)\,\tau^{a}_{(A)}(t')\right]}{\sqrt{(2\pi)^{\mathrm{dim}\mathrm{K}}
\det||\mathrm{K}||}} 
\end{equation}
where $ \sigma_{a(A)}$ represents the rms amplitude $\tau^{a}_{(A)}$ from the 
$a$th source and $\lambda_{a}$ is the characteristic time-scale of the waveform. 
Within $\lambda_{a}$, $\tau^{a}_{(A)}$ is expected to cross the zero level only 
once \cite{adler:1981:tgo}. Therefore, for gravitational wave burst sources such 
as encounters of two supermassive black holes, $\lambda_{a}$ characterizes the 
duration of the burst \cite{deng:2014:sfg}; and for gravitational waves from 
gravity-bound binaries, $\lambda_{a}$ characterizes periods of the binaries. 
\end{subequations}

Eq.~\eqref{eq:tauk} shows that $\tau_{\hat{k}}$ is the linear superposition of 
$\tau^a_{(A)}$ and we assume the two polarization components of the 
gravitational wave from a single source are independent; correspondingly, the 
kernel for the Gaussian process prior of $\tau_{\hat{k}}$ would be the linear 
superposition of Eq.~\eqref{eq:kernelS},
\begin{subequations}
\begin{align} \label{eq:kernelk1}
\mathrm{K}_{\hat{k}\,(jk)}(\Delta t) &= 
\sum_{A}\,F_{j}^{(A)}(\hat{k})F_{k}^{(A)}(\hat{k})\,\sum_{a\,(\hat{k})}\sigma^{2
}_{a(A)}\exp\left(-\frac{\Delta t^2}{2\,\lambda_{a}^2}\right)
\end{align}
where $j,\,k$ denote pulsar indices and $a\,(\hat{k})$ denotes that the sum is 
over all sources in sky location $\hat{k}$. If we assume that at the sky 
location $\hat{k}$, total square sum of all source rms amplitude $\sigma_{a(A)}$ 
is $\sigma^2_{\hat{k}(A)}$ and the density of the sources with square 
characteristic time-scale $\lambda^2_{\hat{k}}$ is 
$\pi_{\hat{k}(A)}\left(\lambda^2_k\right)$, we can approximate the second sum in 
Eq.~\eqref{eq:kernelk1} as an integral over all possible $\lambda^2_{\hat{k}}$, 
i.e.,
\begin{equation}
\sum_{a\,(\hat{k})}\sigma^{2}_{a(A)}\exp\left(-\frac{\Delta 
t^2}{2\,\lambda_{a}^2}\right) = \sigma^{2}_{\hat{k}(A)}\, 
\int\mathrm{d}\lambda^2_{\hat{k}}\,\,\pi_{\hat{k}(A)}(\lambda^{2}_{\hat{k}})\,
\exp\left(-\frac{\Delta t^2}{2\,\lambda^{2}_{\hat{k}}}\right)
\end{equation} 
because the distribution of polarization angle is expected to be uniform 
\cite{misner:1973:g, christensen:1992:mts, flanagan:1993:sot}, so 
$\sigma^{2}_{\hat{k}(A)}$ and $\pi_{\hat{k}(A)}$ are the same for the two GW 
polarization components, i.e.,
\begin{align}
\sigma^{2}_{\hat{k}(+)} = \sigma^{2}_{\hat{k}(\times)} = \sigma^{2}_{\hat{k}}  
\\
\pi_{\hat{k}(+)} = \pi_{\hat{k}(\times)} = \pi_{\hat{k}} 
\end{align}
as a result, the kernel of the Gaussian process prior of $\tau_{\hat{k}}$ 
becomes 
\begin{equation} \label{eq:kernelk0}
\mathrm{K}_{\hat{k}\,(jk)}(\Delta t) = 
\left(\sum_{A}\,F_{j}^{(A)}(\hat{k})F_{k}^{(A)}(\hat{k})\right)\,\sigma^{2}_{
\hat{k}}\,\int\mathrm{d}\lambda^2_{\hat{k}}\,\,\pi_{\hat{k}}(\lambda^{2}_{\hat{k
}})\,\exp\left(-\frac{\Delta t^2}{2\,\lambda^{2}_{\hat{k}}}\right)
\end{equation}
\end{subequations}

Because $\tau^{a}_{(A)}$ is expected to be a smooth function of time, and 
$\tau_{\hat{k}}$ is a linear superposition of a finite number of 
$\tau^{a}_{(A)}$, so $\tau_{\hat{k}}$ should also be a smooth function of time. 
Correspondingly, the kernel of its Gaussian process prior, i.e., 
Eq.~\eqref{eq:kernelk0}, should be infinitely differentiable at $\Delta t = 0$ 
\cite{adler:1981:tgo}. This requirement sets a strong constraint on the choices 
of $\pi_{\hat{k}}$ and the one with the least number of parameters among the 
only few options is that $\pi_{\hat{k}}$ is an inverse gamma distribution of the 
square characteristic time-scale \cite{stein:1999:ios, rasmussen:2006:gpf},
\begin{equation} \label{eq:pi}
 \pi_{\hat{k}}(\lambda^{2}_{\hat{k}}) = 
\frac{\beta_{\hat{k}}^{\alpha_{\hat{k}}}}{\Gamma(\alpha_{\hat{k}})} 
\lambda^{-2\alpha_{\hat{k}}-2}_{\hat{k}} 
\exp\left(-\frac{\beta_{\hat{k}}}{\lambda^{2}_{\hat{k}}}\right)
\end{equation}
where $\alpha_{\hat{k}}$ and $\beta_{\hat{k}}$ are respectively referred to as 
shape parameter and the scale parameter, and they both have to be positive to 
guarantee $\pi_{\hat{k}}$ is normalizable \cite{bernardo:1994:bt, 
gelman:2004:bda}; $\Gamma(\alpha_{\hat{k}})$ is the gamma function of 
$\alpha_{\hat{k}}$ \cite{abramowitz:1964:mcs}. Combining Eq.~\eqref{eq:kernelk0} 
and Eq.~\eqref{eq:pi}, we can obtain the kernel of the Gaussian process prior of 
$\tau_{\hat{k}}$
\begin{equation}\label{eq:kernelk}
 \mathrm{K}_{\hat{k}\,(jk)}(\Delta t) = \sigma^{2}_{\hat{k}} 
\left(1+\frac{\Delta 
t^2}{2\alpha_{\hat{k}}\xi_{\hat{k}}^2}\right)^{-\alpha_{\hat{k}}}\, 
\sum_{A}\,F_{j}^{(A)}(\hat{k})F_{k}^{(A)}(\hat{k})
\end{equation}
where $\xi_{\hat{k}} = \sqrt{\beta_{\hat{k}}}$, which is the characteristic 
time-scale of $\tau_{\hat{k}}$. 

Because the gravitational wave background is the superposition of the 
gravitational waves from all directions in the sky, so kernel of the Gaussian 
process prior of the timing residuals induced by the background should be the 
sum of Eq.~\eqref{eq:kernelk} across the whole sky. %However, as we discussed 
above, pulsar timing arrays can only localize the gravitational wave sources 
$\gtrsim 1000\,\mathrm{deg^2}$, which covers a few percent of the sky; 
correspondingly, we expect that the pulsar timing arrays should not be sensitive 
to the anisotropy of the gravitational wave background. We will demonstrate this 
point by representative examples in Sec.~\ref{sec:demo}. 
In general, the gravitational wave sources may not be isotropically distributed 
and the anisotropy of the background can be characterized by decomposing the 
angular distribution of the gravitational wave energy density on the sky into 
multipole moments \cite{mingarelli:2013:cgw, taylor:2013:sfa}. However, for the 
purpose of demonstration, in this paper we only focus on the isotropic 
gravitational wave background and it is straightforward to generalize our method 
to anisotropic cases by combining the techniques presented in 
\cite{mingarelli:2013:cgw, taylor:2013:sfa}. By assuming isotropy, 
$\tau_{\hat{k}}$ in all directions will have the same $\sigma_{\hat{k}}$, 
$\alpha_{\hat{k}}$ and $\xi_{\hat{k}}$. Correspondingly, the kernel of the 
gravitational wave background induced timing residuals $\bm{\tau}$, i.e., 
Eq.~\eqref{eq:kernel0}, is 
\begin{subequations} \label{eq:kernel}
\begin{equation}
 \mathrm{K}_{jk}(\Delta t) = \sigma^{2} \left(1+\frac{\Delta 
t^2}{{2\alpha}\xi^2}\right)^{-\alpha} \, \gamma_{jk}
\end{equation}
where $\gamma_{jk}$ is
\begin{align}
 \gamma_{jk} &= \int\mathrm{d^2}\Omega_{\hat{k}} \, 
\sum_{A}\,F_{j}^{(A)}(\hat{k})F_{k}^{(A)}(\hat{k}) \nonumber \\
&= 
\frac{3}{2}\frac{1-\hat{n}_j\cdot\hat{n}_k}{2}\log\left(\frac{1-\hat{n}
_j\cdot\hat{n}_k}{2}\right) -\frac{1}{4}\frac{1-\hat{n}_j\cdot\hat{n}_k}{2} + 
\frac{1}{2} + \frac{1}{2} \delta_{jk}
\end{align}
\end{subequations}
We can see that $\gamma_{jk}$ is the Hellings-Downs Curve 
\cite{hellings:1983:ulo}. 

Combining Eq.~\eqref{eq:prior0} and Eq.~\eqref{eq:kernel}, we can obtain the 
Gaussian process prior of $\bm{\tau}$
\begin{equation} \label{eq:priortau}
 q(\bm{\tau}|\sigma,\xi,\alpha) = 
\frac{\exp\left[-\frac{1}{2}\bm{\tau}^{T}\bm{\mathrm{K}}^{-1}\bm{\tau}\right]}{
\sqrt{(2\pi)^{\mathrm{dim}\bm{\mathrm{K}}}\det||\bm{\mathrm{K}}||}}
\end{equation}
where $\bm{\mathrm{K}}$ is expressed by Eq.~\eqref{eq:kernel}

\subsubsection{Prior of Hyperparameters}
The prior probability density of $\bm{\tau}$ Eq.~\eqref{eq:priortau} includes 
three undermined hyperparameters --- $\sigma$, $\xi$ and $\alpha$, and we need 
to choose their prior probability density to make a full Bayesian inference. 

For $\sigma$, it is a scale factor and we can choose Jeffreys prior 
\cite{jeffreys:1946:aif}. However, Jeffreys prior will make the posterior 
probability density improper \cite{deng:2014:sfg, gelman:2006:pdf} and 
correspondingly such an uninformative prior is not an appropriate one. In this 
case, a uniform distribution from $0$ to $+\infty$, which will make the 
posterior distribution normalizable, is recommended \cite{deng:2014:sfg, 
gelman:2006:pdf}, i.e.,
\begin{subequations} \label{eq:hyperprior}
\begin{equation}
q_\sigma(\sigma) \propto 1
\end{equation}

For $\xi$, it is a time-scale factor and choose the Jeffreys prior
\begin{equation}
 q_\xi(\xi) \propto 1/\xi
\end{equation}

The hyperparameter $\alpha$ is the shape parameter of the inverse gamma 
distribution Eq.~\eqref{eq:pi}, which represents the number density of the 
sources. We expect that the number of sources should monotonically increase with 
the increase of their periods or durations \cite{sesana:2008:tsg}, 
correspondingly, $\pi_{\hat{k}}$ in Eq.~\eqref{eq:pi} should be a monotonically 
increasing function of $\lambda_{\hat{k}}$. To satisfy this requirement, 
$\alpha$ should be between $0$ and $1$ \cite{gelman:2004:bda, bernardo:1994:bt}. 
Since $\alpha$ is a dimensionless hyperparameter, the corresponding 
uninformative prior should be a uniform distribution between $0$ and $1$ 
\cite{bernardo:1994:bt},
\begin{equation} 
 q_\alpha(\alpha) = 1
\end{equation}
 \end{subequations}

\subsection{Bayesian Nonparametric Inferences} \label{sec:inference}
\subsubsection{Inferring $\boldsymbol{\tau}$ and hyperparameters} 
Since we have obtained the likelihood function and chosen the appropriate priors 
for $\bm{\tau}$ and hyperparameters, we can 
follow the discussion in Sec.~\ref{sec:frame} and make the inference of 
$\bm{\tau}$ and hyperparameters. Combining Eq.~\eqref{eq:postertau} 
with Eq.~\eqref{eq:likelihood}, Eq.~\eqref{eq:priortau} and 
Eq.~\eqref{eq:hyperprior}, we can determine the joint posterior 
probability density of $\bm{\tau}$ and hyperparameters,
\begin{subequations}\label{eq:inference}
\begin{align} \label{eq:posterior}
p(\bm{\tau},\sigma, \xi, \alpha|\bm{\mathrm{d}}) &= \frac{1}{Z(\bm{\mathrm{d}})} 
\Lambda(\bm{\mathrm{d}}|\bm{\tau})q(\bm{\tau}|\sigma,\xi,
\alpha)q_\sigma(\sigma)q_\xi(\xi)q_\alpha(\alpha) \nonumber \\
&= 
\sqrt{\frac{\det||\bm{\mathrm{A}}||}{(2\pi)^{\mathrm{dim}\bm{\mathrm{A}}}}}
\exp\left[-\frac{1}{2}(\bm{\tau}-\bm{\tau}_{m})^{T}\bm{\mathrm{A}}(\bm{\tau}-\bm
{\tau}_{m})\right] \nonumber \\
&\times 
\frac{1}{Z(\bm{\mathrm{d}})}\Lambda_\theta(\bm{\mathrm{d}}|\sigma,\xi,
\alpha)q_\sigma(\sigma)q_\xi(\xi)q_\alpha(\alpha)
\end{align}
where $\bm{\mathrm{A}}$ is 
\begin{equation}
\bm{\mathrm{A}} = \bm{\mathrm{K}}^{-1} + \bm{\mathrm{C}}^{-1}
\end{equation}
and $\bm{\tau}_m$ satisfies
\begin{equation}
\bm{\mathrm{A}} \bm{\tau}_m = \bm{\mathrm{C}}^{-1} \bm{\mathrm{d}}
\end{equation}
$\Lambda_\theta(\bm{\mathrm{d}}|\sigma, \xi, \alpha)$ is the likelihood function 
of hyperparameters after marginalizing over $\bm{\tau}$:
\begin{align}\label{eq:marginlike} 
\Lambda_\theta(\bm{\mathrm{d}}|\sigma, \xi, \alpha) &= \int 
\Lambda(\bm{\mathrm{d}}|\bm{\tau})q(\bm{\tau}|\sigma, \xi, \alpha) \, 
\mathrm{d}\bm{\tau} \nonumber \\
&= 
\frac{\exp\left[-\frac{1}{2}\bm{\mathrm{d}}^{T}\bm{\mathrm{C}}^{-1}\bm{\mathrm{d
}}\right]}{\sqrt{(2\pi)^{\mathrm{dim}\,\bm{\mathrm{d}}}\det||\bm{\mathrm{C}}||}} 
\times 
\frac{\exp\left[\frac{1}{2}(\bm{\mathrm{C}}^{-1}\bm{\mathrm{d}})^{T}\bm{\mathrm{
A}}^{-1}(\bm{\mathrm{C}}^{-1}\bm{\mathrm{d}})\right]}{\sqrt{\det||\bm{\mathrm{A}
}||\det||\bm{\mathrm{K}}||}} \nonumber \\
&= 
\frac{\exp\left[-\frac{1}{2}\bm{\mathrm{d}}^{T}(\bm{\mathrm{C}}+\bm{\mathrm{K}}
)^{-1}\bm{\mathrm{d}}\right]}{\sqrt{(2\pi)^{\mathrm{dim}\,\bm{\mathrm{d}}}
\det||\bm{\mathrm{C}}+\bm{\mathrm{K}}||}}
\end{align}
and the last row of Eq.~\eqref{eq:posterior} becomes the marginal posterior probability density of the hyperparameters:
\begin{equation}\label{eq:mp}
 p(\sigma, \xi, \alpha|\bm{\mathrm{d}}) = \frac{1}{Z(\bm{\mathrm{d}})}\Lambda_\theta(\bm{\mathrm{d}}|\sigma,\xi,
\alpha)q_\sigma(\sigma)q_\xi(\xi)q_\alpha(\alpha)
\end{equation}

Therefore, the Bayesian nonparametric analysis described above for detecting a 
gravitational wave background is identical to the Bayesian hierarchical modeling
(e.g.~\cite{lentati:2013:hmb}). We can see that the marginal likelihood function in Eq.~\eqref{eq:marginlike} 
coincides the one used in previous methods that assume the gravitational wave 
background is generated from a Gaussian process with covariance matrix 
$\bm{\mathrm{K}}$. However, at the end of this section, we will discuss that 
different interpretations of the marginal likelihood in 
Eq.~\eqref{eq:marginlike} will lead to different choices of $\bm{\mathrm{K}}$.
\end{subequations}

Eq.~\eqref{eq:inference} summarizes Bayesian nonparametric inference, which 
gives estimation on $\bm{\tau}$ and hyperparameters. 

\subsubsection{Inferring the Presence of A Gravitational Wave Background} 
\label{sec:DIC}
Given timing residual observations $\bm{\mathrm{d}}$ from an array of pulsars, 
we would like to infer if a gravitational wave background is present. We treat 
this question as a problem in Bayesian model comparison \cite{gelman:2004:bda}. 
Consider the two models
\begin{subequations}\label{eq:model}
\begin{align} 
&M_{1} = \left(\text{a gravitational wave background is present in the data 
set}\right) \\
&M_{0} = \left(\text{no gravitational waves backgrounds are present in the data 
set}\right)  
\end{align}
\end{subequations}
The purpose of model comparison is to check which model data favors. If data 
favors $M_1$, then it indicates that a 
gravitational wave is likely to be present in the data set. Bayes factor is 
often used for the criterion of this model comparison problem. However, if the 
prior distribution of the parameters is improper, there would be an arbitrary 
multiplicative factor in Bayes factor, which makes it ill-defined 
\cite{kass:1995:bf, gelman:2004:bda, deng:2014:sfg}. Correspondingly, Bayes 
factor may not be a suitable criterion for our case, because the prior 
distributions of RMS amplitude $\sigma$ and the characteristic time-scale $\xi$ 
are improper. Therefore, we use an alternative criterion {\it Deviance 
Information Criterion} (DIC) \cite{spiegelhalter:2002:bmo}, which is the sum of 
two terms --- one term represents ``goodness of fitting'', which 
measures how well the model fits the data; the other term represents ``the 
penalty of complexity'', which measures how complex the model is 
\cite{claeskens:2008:msa}. Here we briefly summarize the principle of DIC. We 
refer readers to \cite{spiegelhalter:2002:bmo} for detail discussion in general 
and \cite{deng:2014:sfg} for application in gravitational wave context. 
\cite{deng:2014:sfg} also discussed why DIC is more applicable than the commonly 
used Bayes factor in gravitational wave context. 

In DIC, the ``goodness of fitting'' is summarized in ``deviance'', defined as 
$-2$ times the log-likelihood \cite{gelman:2004:bda}:
\begin{equation}
 D(\bm{\mathrm{d}},\bm{\tau}) = -2 \log \Lambda(\bm{\mathrm{d}}|\bm{\tau})
\end{equation}
which measures the model discrepancy and resembles the classical $\chi^2$ 
goodness-of-fit measure. The average of 
the deviance on posterior probability distribution provides a summary of the 
error of model $M_1$ and represents ``goodness of fitting'' 
\cite{spiegelhalter:2002:bmo}:
\begin{subequations}
\begin{equation} \label{eq:deviance}
D_{\mathrm{avg}}(\bm{\mathrm{d}},M_1) = \int D(\bm{\mathrm{d}},\bm{\tau}) 
p_\tau(\bm{\tau}|\bm{\mathrm{d}}) \mathrm{d}\bm{\tau}
\end{equation}
where $p_\tau(\bm{\tau}|\bm{\mathrm{d}})$ is the posterior probability density 
in Eq.~\eqref{eq:posterior} marginalizing over all hyperparameters:
\begin{equation} \label{eq:posteriortau}
p_\tau(\bm{\tau}|\bm{\mathrm{d}}) = \int \, 
p(\bm{\tau},\sigma,\xi,\alpha|\bm{\mathrm{d}})\, 
\mathrm{d}\sigma\mathrm{d}\xi\mathrm{d}\alpha
\end{equation}
\end{subequations}

For model $M_0$, since there are no parameters representing the model, the 
average of the deviance is 
\begin{subequations}
\begin{equation}
D_{\mathrm{avg}}(\bm{\mathrm{d}},M_0) = -2 \log \Lambda(\bm{\mathrm{d}}|M_0)
\end{equation}
where $\Lambda(\bm{\mathrm{d}}|M_0)$ is the null model likelihood function
\begin{equation} \label{eq:deviance0}
\Lambda(\bm{\mathrm{d}}|M_{0}) = N(\bm{\mathrm{d}}|\bm{\mathrm{C}}) = 
\frac{\exp\big(-\frac{1}{2}\bm{\mathrm{d}}^{T}\bm{\mathrm{C}}^{-1}\bm{\mathrm{d}
}\big)}{\sqrt{(2\pi)^{\mathrm{dim}\,\bm{\mathrm{d}}}\det||\bm{\mathrm{C}}||}}
\end{equation}
\end{subequations} 

Now we need to consider the complexity of a model and the more complex model 
with more adjustable parameters should have larger 
penalty \cite{gelman:2004:bda, claeskens:2008:msa, jefferys:1991:sor}. The 
complexity of a model is represented by the measure of the degree of 
overfitting. In DIC, it is defined as the difference between the posterior mean 
deviance Eq.~\eqref{eq:deviance} 
and the deviance at the mean value of $\bm{\tau}$ under its posterior 
probability distribution Eq.~\eqref{eq:posteriortau} for 
model $M_1$ \cite{spiegelhalter:2002:bmo}
\begin{subequations}
\begin{equation} \label{eq:complexity}
p_{D}(\bm{\mathrm{d}},M_1) = D_{\mathrm{avg}}(\bm{\mathrm{d}},M_1) - 
D(\bm{\mathrm{d}},\bar{\bm{\tau}})
\end{equation}
where $\bar{\bm{\tau}}$ is the mean value of $\bm{\tau}$ under its posterior 
probability distribution Eq.~\eqref{eq:posteriortau},
\begin{equation} \label{eq:meantau}
\bar{\bm{\tau}} = \int \, \bm{\tau} \, p_\tau(\bm{\tau}|\bm{\mathrm{d}}) \, 
\mathrm{d}\bm{\tau}
\end{equation}
\end{subequations}
$p_D$ can be thought as the reduction in the lack of fit due to Bayesian 
estimation, or alternatively the degree of overfitting due to $\bar{\bm{\tau}}$ 
adapting to the data set $\bm{\mathrm{d}}$ \cite{spiegelhalter:2002:bmo}, since 
$\bar{\bm{\tau}}$ serves as a Bayesian estimator of the model, and 
correspondingly $D(\bm{\mathrm{d}},\bar{\bm{\tau}})$ represents the lack of fit 
to the data due to the Bayesian estimation of the model. 

For model $M_0$, since no parameters or functions are needed, so the effective 
number of parameters for this model, $p_{D}(\bm{\mathrm{d}},M_0)$, is zero.

The sum of the average of the deviance and the effective number of parameters 
$p_{D}$ is referred to as Deviance Information Criterion 
\cite{spiegelhalter:2002:bmo},
\begin{equation} \label{eq:DIC}
\mathrm{DIC}(\bm{\mathrm{d}},M) =  D_{\mathrm{avg}}(\bm{\mathrm{d}},M) + 
p_{D}(\bm{\mathrm{d}},M)
\end{equation}
The data favors the model with smaller DIC, since such a model has smaller 
discrepancy of the data and is less complex.

The difference between the DICs of two models in Eq.~\eqref{eq:model} ,
\begin{equation}
\Delta \mathrm{DIC} = \mathrm{DIC}(\bm{\mathrm{d}},M_{1}) - 
\mathrm{DIC}(\bm{\mathrm{d}},M_{0})
\end{equation}
characterizes the relative odds between the two models. Therefore, the 
difference of DICs between two models is similar to 
likelihood ratio test statistic \cite{neyman:1933:otp} and twice the natural 
logarithm of Bayes factor \cite{kass:1995:bf}. Correspondingly, it has the same 
scale as those statistics \cite{spiegelhalter:2002:bmo}. If $\Delta \mathrm{DIC} 
\lesssim -10$, it is safe to conclude that the data 
strongly favors $M_{1}$ and there is strong evidence that a gravitational wave 
background is present in the data set \cite{spiegelhalter:2002:bmo}.

\subsection{Discussion of Deterministic and Stochastic Modeling}
At this point, it is worth comparing the deterministic modeling by Bayesian 
nonparametrics discussed above and the stochastic modeling used by previous 
methods. 

In general, when we detect a signal across a finite time duration, we have two 
ways to model the signal:
\begin{itemize}
\item[(1)] assume the signal is generated by a deterministic process. However, 
we do not know the function form of the signal, on which we need to assign a 
prior distribution. This is what we did in this paper.
\item[(2)] assume the signal is a random sample (one single realization) 
generated from a stochastic process, and what we need to do is to model the 
distribution function of the stochastic process. This is what previous methods 
did. 
\end{itemize}
Because we do not have an ensemble of the signals and we cannot reverse time to 
repeat the detection, so both of these modeling methods may lead to reasonable 
characterizations of the signal. Which is more effective depends on which model 
fits the data better, i.e., which method leads to a larger likelihood or a 
smaller DIC. 

In the case of detecting a gravitational wave background, both of these two 
methods would result in the same marginal likelihood function 
Eq.~\eqref{eq:marginlike}, because in the first method, we assume the prior 
distribution of the signal form is a Gaussian distribution, and in the second 
method, we assume that the distribution function of the gravitational wave 
background is also a Gaussian distribution. This means that both of these two 
methods may lead to the same inference of the signal. So in practice, the method
we present above may be considered as the same method as previous ones in \cite{haasteren:2009:omt, lentati:2013:hmb} except
the difference choices of the kernel $\bm{\mathrm{K}}$. However, the choices of 
the kernels strongly depend on what logic we follow, which will lead to different 
values of likelihood functions. This is the key difference between the method presented here
and the previous methods. 

When we follow the first approach, as described in Section \ref{sec:tauprior}, the 
kernel $\bm{\mathrm{K}}$, which originally appears in the prior distribution, is 
chosen to characterize the expected shape of the signal, such as its smoothness, 
its variation, its trend, etc. We use these characteristics to represent the 
underlying unknown deterministic function form of the signal. Following this 
logic, we finally obtain the appropriate kernel Eq.~\eqref{eq:kernel}. While if 
we follow the second approach, $\bm{\mathrm{K}}$ characterizes the covariance of the 
Gaussian distribution, which is assumed to be the underlying distribution 
function of the gravitational wave background. Following this logic, 
$\bm{\mathrm{K}}$ would have to be the Fourier transform of a frequency power 
law, because it can be derived from physics that the frequency distribution of 
the gravitational wave sources follows a power law \cite{haasteren:2009:omt, 
lentati:2013:hmb}. Using the first method cannot lead to the choice of a power 
law spectrum while using the second method cannot lead to the choice of 
Eq.~\eqref{eq:kernel}. The two different choices of $\bm{\mathrm{K}}$ would lead 
to different values of the likelihood functions.

If the non-Gaussian part of the distribution of the gravitational wave 
background is significant, the second method will be ineffective because it only 
characterizes the covariance of the distribution but ignores the skewness, 
kurtosis and other parts of the distribution. We may wonder how the first method 
can be applied here. The first method does not try to characterize the 
underlying distribution of the gravitational wave background. The signal we 
detect is only one single sample (one single realization) generated from the 
underlying distribution, and without an ensemble of the signals, we may be 
hardly able to characterize the underlying distribution. Instead, the first 
method assumes the signal is just a representation of some deterministic 
function but does not worry if there is some underlying distribution. However, 
since we do not know the exact function form of the signal, we assign a prior 
distribution with a specific kernel to characterize the expected shape of the 
signal. Correspondingly, no matter if the signal is sampled from some 
distribution or what the distribution is, as long as our Gaussian prior with a 
kernel correctly characterizes the shape of the signal pattern, such as its 
smoothness, variation, trend, etc., our method would lead to a good inference of 
the signal. 

To recap, even though for detecting a gravitational wave background, the 
deterministic modeling by our Bayesian nonparametrics and the previous 
stochastic modeling will lead to likelihood functions with the same form, our 
method only tries to characterize the signal itself and the kernel in the 
Gaussian prior distribution represents our expectation of the shape of the 
signal pattern; while previous methods assume the signal is sampled from some 
underlying distribution and try to use a Gaussian model with a power-law 
spectrum to characterize the distribution. These two approaches will lead to 
different choices of the kernels, which correspondingly would result in 
different values of likelihood functions.

% flatex input end: [bayes.tex]

%\usepackage{hyperref}

% flatex input: [demo.tex]
\section{Examples} \label{sec:demo}
\subsection{Overview}
To illustrate the effectiveness of the analysis techniques described above, we 
apply them to simulated data sets of 4 millisecond pulsars in the current 
International Pulsar Timing Array (IPTA) \cite{hobbs:2010:tip, 
demorest:2012:lot, manchester:2012:tpp} which are most accurately timed as 
described in Table~\ref{tab:IPTA}. The capability of detecting and 
characterizing gravitational waves is dominated by these best pulsars, although 
they are the minority of the full IPTA \cite{burt:2011:opt}. We will also 
compare our method to the conventional one proposed by van Haasteren et al 
\cite{haasteren:2009:omt}, which assumes the background is exactly Gaussian . 

\begin{table}[ht]
\centering
\caption[4 IPTA pulsars used to detect a gravitational wave 
background]{\hspace{0.3cm} 4 IPTA pulsars we use, Their white timing noise rms 
and the Telescopes from which the timing residuals are measured 
\cite{hobbs:2010:tip, demorest:2012:lot, manchester:2012:tpp}}
\vspace{0.5cm}
\begin{tabular}{c c c}
\hline \hline
Pulsar \hspace{1.5cm} & rms (ns) & \hspace{1.5cm} Telescope \\
\hline
J1713$+$0747 \hspace{1.5cm} & 30 &  \hspace{1.5cm} AO \\
J1909$-$3744 \hspace{1.5cm} & 38 & \hspace{1.5cm} GBT \\
J0437$-$4715 \hspace{1.5cm} & 75 & \hspace{1.5cm} Parkes \\
J1857$+$0943 \hspace{1.5cm} & 111 & \hspace{1.5cm} AO \\
\hline
\end{tabular}
\label{tab:IPTA}
\end{table}

We uniformly sample $50$ observation times across $5$ year observation, and the 
corresponding pulsar timing data sets are constructed by 
\begin{itemize}
\item[(1)] evaluating the pulsar timing residuals induced by a simulated 
gravitational wave background that will be described in Sec.~\ref{sec:signal}.
\item[(2)] adding pulsar timing noise that will be described in 
Sec.~\ref{sec:noise} to the timing residuals obtained by the first step.
\item[(3)] removing the linear trend of the timing residuals obtained above to 
simulate the procedure in the standard pulsar timing analysis that removes the 
effects of pulsar spin and spin down. 
\end{itemize}
When we analyze the data, we add a linear model in the pulsar timing response 
function to account for the linear trend, the same as the analysis in 
\cite{haasteren:2009:omt}. We will also apply our analysis methods to a data set 
composed of timing noise alone for comparative study. 

\subsection{Construction of Simulated data Sets}
\subsubsection{Simulated Gravitational Wave Background} \label{sec:signal}
We construct the simulated observations of two isotropic gravitational wave 
backgrounds respectively generated from $10^6$ and $10^5$ supermassive black 
hole binaries. Both of these sources are generated in the same way as we did in 
Section \ref{sec:nonGaussian}:
\begin{itemize}
\item[(1)] the frequency distribution of these sources follows 
Eq.~\eqref{eq:density} with lower bound of $0.2\,\mathrm{yr}^{-1}$ and upper 
bound of $4\,\mathrm{yr}^{-1}$.
\item[(2)] all of these sources are uniformly distributed across the sky with 
their gravitational wave peak timing residuals ranging from $0.01\mathrm{ns}$ to 
$100\mathrm{ns}$.
\item[(3)] the orbital orientations and initial phases of all these sources are 
uniformly distributed.
\item[(4)] the timing residuals induced by the gravitational wave background are 
the sum of all the gravitational wave signals generated from the sources sampled 
from the distribution described in the above three steps. The RMS values of the 
gravitational wave amplitudes of the two gravitational wave backgrounds are both 
about $22\mathrm{ns}$.
\end{itemize}

The degrees of non-Gaussianity of these two backgrounds are presented in 
Table~\ref{tab:ngauss}, and we can see the background with $10^5$ sources is 
more non-Gaussian than that with $10^6$ sources. For these two cases of 
simulated backgrounds, we will compare the results of our method and the 
conventional one proposed by van Haasteren et al \cite{haasteren:2009:omt} and 
we will see that our method is much more effective on the case of the background 
with $10^5$ sources.

\subsubsection{Pulsar Timing Noise} \label{sec:noise}
The millisecond pulsars used in current International pulsar timing array 
typically show white noise on short timescales, and 
few of them turn to red noise on timescales $\gtrsim 5$ years 
\cite{demorest:2012:lot, manchester:2012:tpp}. For demonstrations, we use the 
noise model described in \cite{deng:2014:sfg} and we briefly summarize it here. 
The power spectral density is taken to be \cite{coles:2011:pta}
\begin{subequations}
\begin{equation}\label{eq:noise}
S_{n}(f) = 
\sigma^{2}_{n}+\sigma^{2}_{n}\left[\frac{1+\left(\frac{f}{f_0}\right)^2}{
1+\left(\frac{f_r}{f_0}\right)^2}\right]^{-5/2}
\end{equation}
where
\begin{align}
\sigma_{n} &= \left(\text{white noise rms}\right) \\
f_r & = \left(\text{red-white noise cross-over frequency, 
$0.2\,\mathrm{yr^{-1}}$}\right)
\end{align}
\end{subequations}
and $f_0$ softens the noise spectrum at ultra-low frequency. As long as $f_0$ is 
much less than the pulsar timing array frequency 
band, its value does not matter. In the simulation we set $f_0$ equal to 
$0.01\mathrm{nHz}$. We choose the power index of the red noise 
spectrum as $-5$ because the few millisecond pulsars showing red noise have 
noise spectrum with power index $-5$ \cite{shannon:2010:atr}. 

The covariance matrix of the noise will be the fourier transform of the noise 
specturm density Eq.~\eqref{eq:noise} to time 
domain, i.e., 
\begin{equation} \label{eq:cov}
\mathrm{C}(t_i,\, t_j) = \sigma^{2}_{n}\left(\delta_{ij} + \sqrt{\frac{2}{9\pi}} 
\left[1+\left(\frac{f_r}{f_0}\right)^2\right]^{5/2}f^3_0|t_i-t_j|^{2}K_{2}
(f_0|t_i-t_j|)\right)
\end{equation}
where $t_{i,j}$ are the ``observation times'' and $K_2$ is the modified Bessel 
function of the second kind with index $2$. The pulsar 
timing noise for each pulsar are sampled from multivariate normal distribution 
with zero mean and covariance matrix Eq.~\eqref{eq:cov}.

\subsection{Analysis of Simulated Data Sets}
Our Bayesian nonparametric analysis is designed to investigate if a 
gravitational wave background is present in the dataset, and also infer the 
hyperparameters. We use Metropolis-Hasting method of Markov Chain Monte Carlo \cite{robert:2004:mcs} 
to compute the posterior probability densities and Deviance Information 
Criterion described in Sec.~\ref{sec:inference}. We follow the same computing procedure
as in \cite{haasteren:2009:omt} to sample the posterior probability distribution Eq.~\eqref{eq:mp} of 
the three hyperparameters, except we use the Cauchy distribution as the proposal distribution. We sample $10^6$ 
step random walks by Metropolis algorithm and it takes about three hours for analysis of each data set described above. 

We apply both our method and 
the conventional Gaussian method on the two data sets --- one contains the 
contribution of the gravitational wave background with $10^6$ sources and the 
other contains the contribution of the background with $10^5$ sources. We also 
analyze the data set consisting of timing noise alone only by our method for 
comparative study. Table~\ref{tab:results} lists the results of the analysis. 
The parentheses in the second column contain the DIC differences obtained by 
applying conventional Gaussian method in \cite{haasteren:2009:omt}.

\begin{table}[ht]
\centering
\caption{\hspace{0.1cm} Results for Bayesian Nonparametric Analysis on 3 
Simulated Data Sets}
\vspace{0.5cm} 
\begin{tabular}{c c c c c}
\hline \hline
No. of Sources \hspace{1cm} & $\Delta\mathrm{DIC}$ (Gaussian) \hspace{1cm} & 
$\epsilon_{\sigma}$  \hspace{1cm} & $\epsilon_{\xi}$ \hspace{1cm} & 
$\epsilon_{\alpha}$\\
\hline
$10^6$  \hspace{1cm} & -15 (-12) \hspace{1cm} & $27.4\%$ \hspace{1cm} & $96.8\%$ 
\hspace{1cm} & $32.0\%$\\
$10^5$  \hspace{1cm} & -14 (-3) \hspace{1cm} & $29.2\%$  \hspace{1cm} &  
$71.9\%$ \hspace{1cm} & $34.2\%$\\
Absent  \hspace{1cm} & 5 (5)  \hspace{1cm} & $92.5\%$  \hspace{1cm} &  $63.5\%$ 
\hspace{1cm} & $54.6\%$\\
\hline
\end{tabular}
\begin{flushleft}
{\bf Notes.} The signals correspond to an isotropic gravitational wave 
background and an anisotropic one described in Sec.~\ref{sec:signal}. 
$\epsilon_{\sigma}$, $\epsilon_{\xi}$ and $\epsilon_{\alpha}$ respectively 
denote the fractional uncertainty of $\sigma$, $\xi$ and $\alpha$. The 
parentheses contain the DIC differences by applying the conventional Gaussian 
method on the same data sets. 
\end{flushleft}
\label{tab:results}
\end{table}

\subsubsection{Signal of an Isotropic Gravitational Wave Background with $10^6$ 
Sources} \label{sec:isotropic}
We simulate an isotropic gravitational wave background by sampling $10^6$ 
sources from a homogeneous and isotropic distribution Eq.~\eqref{eq:density} and 
computing the superposition of the timing residuals induced by the gravitational 
waves from them, as described in Sec.~\ref{sec:signal}. The first row of 
Table.~\ref{tab:results} and Fig.~\ref{fig:ihyper} summarize the results of our 
Bayesian nonparametric analysis:

\begin{itemize}
\item[(1)] From the first row of Table.~\ref{tab:results}, we see that the 
difference between the DICs of the positive hypothesis and null hypothesis, 
described in Sec.~\ref{sec:DIC}, is $-15$, corresponding to a strong evidence 
for the presence of a gravitational wave background in the data set. We apply 
the conventional Gaussian method in van Haasteren et al 
\cite{haasteren:2009:omt}on the same data set and the DIC difference is $-12$, 
which also indicates a strong evidence of the presence of a gravitational wave 
background. Therefore, the number of sources in this cases is not small enough to
distinguish the effectiveness of the two analysis methods. 
\item[(2)] We also infer the hyperparameters and Fig.~\ref{fig:ihyper} shows the 
posterior probability density of them. The mean value of $\sigma$ is 
$22.5\,\mathrm{ns}$ (consistent with the simulated signal of RMS amplitude 22ns) 
and its rms errors are respectively $6.2\,\mathrm{ns}$. Correspondingly, its 
fractional uncertainty is $27.4\%$. For $\xi$, the mean value is 
$0.25\,\mathrm{yr}$ and the rms error is $0.24\,\mathrm{yr}$, and the 
corresponding fractional uncertainty is $96.8\%$. We cannot measure the shape 
parameter $\alpha$ very well and it tends to be $1$. The mean and rms error are 
respectively $0.69$ and $0.22$, and the fractional error is $32.0\%$. 
Fig.~\ref{fig:igauss} shows the posterior probability density of the strain 
amplitude $\sigma_{\mathrm{GW}}$ and the spectrum power index 
$\alpha_{\mathrm{GW}}$ the gravitational wave background obtained by applying the conventional method in 
\cite{haasteren:2009:omt}.

\begin{figure}[ht]
\centering
\includegraphics[width=15cm]{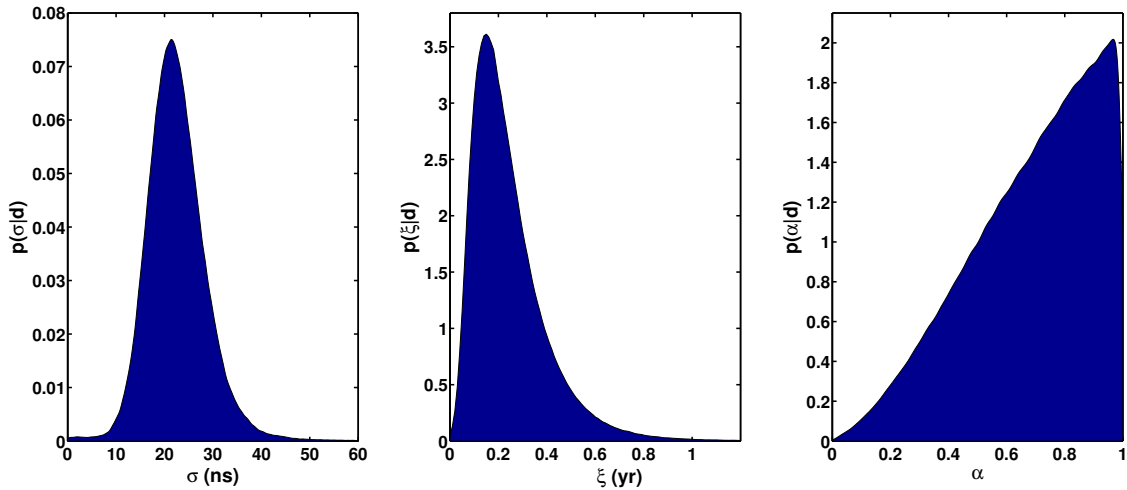}
\caption[Posterior of hyperparameters of the nonparametric model for the 
non-Gaussian gravitational wave background]{Posterior probability densities of 3 
hyperparameters --- $\sigma$, $\xi$ and $\alpha$, for analysis on the data 
described in Sec.~\ref{sec:isotropic}. The fractional errors of them are 
respectively $27.4\%$, $96.8\%$, $32.0\%$}
\label{fig:ihyper}
\end{figure}

\begin{figure}[ht]
\centering
\includegraphics[width=15cm]{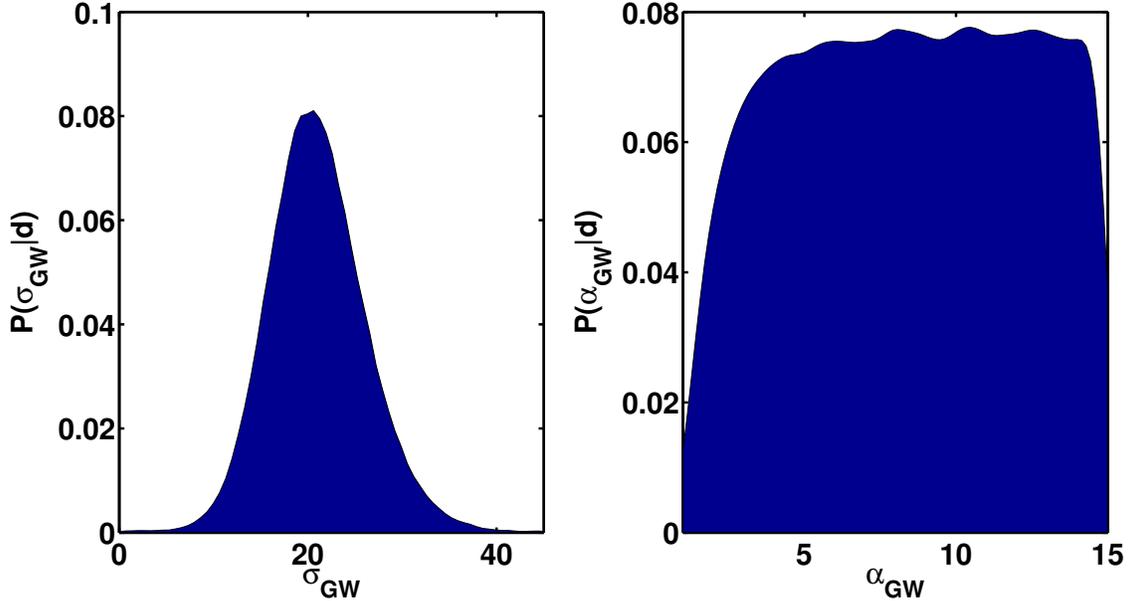}
\caption[Posterior of parameters of the Gaussian parametric model for the 
non-Gaussian background]{Posterior probability densities the strain amplitude 
$\sigma_{\mathrm{GW}}$ and the spectrum power index $\alpha_{\mathrm{GW}}$ the 
gravitational wave background with $10^6$ sources obtained by applying the conventional method in 
\cite{haasteren:2009:omt}.}
\label{fig:igauss}
\end{figure}

\end{itemize}

\subsubsection{Signal of an Isotropic Gravitational Wave Background with $10^5$ 
Sources} \label{sec:anisotropic}
We simulate an isotropic gravitational wave background by sampling $10^5$ 
sources from a homogeneous and isotropic distribution Eq.~\eqref{eq:density} and 
computing the superposition of the timing residuals induced by the gravitational 
waves from them, as described in Sec.~\ref{sec:signal}. The degree of 
non-Gaussianity of this background is greater than the one with $10^6$ sources 
as illustrated in Table~\ref{tab:ngauss}. The second row of 
Table.~\ref{tab:results} and Fig.~\ref{fig:ahyper} summarize the results of our 
Bayesian nonparametric analysis on such ``anisotropic signal'' data:

\begin{itemize}
\item[(1)] From the second row of Table.~\ref{tab:results}, we see that the 
difference between the DICs of the positive hypothesis and null hypothesis, 
described in Sec.~\ref{sec:DIC}, is $-14$, corresponding to a strong evidence 
for the presence of a gravitational wave background in the data set. When we 
apply the conventional Gaussian method in \cite{haasteren:2009:omt} on the data 
set, we obtain a DIC difference of only $-3$, which indicates no strong evidence 
of a gravitational wave background. Therefore, the non-Gaussianity of this 
background is non-negligible, and our method shows the strong advantage over the 
conventional one in this case.

\item[(2)] We also infer the hyperparameters and Fig.~\ref{fig:ahyper} shows the 
posterior probability density of them. The mean value of $\sigma$ is 
$22.4\,\mathrm{ns}$ (consistent with the simulated signal of RMS amplitude 22ns) 
and its rms errors are respectively $6.5\,\mathrm{ns}$. Correspondingly, its 
fractional uncertainty is $30.0\%$. For $\xi$, the mean value is 
$0.36\,\mathrm{yr}$ and the rms error is $0.26\,\mathrm{yr}$, and the 
corresponding fractional uncertainty is $71.9\%$. We cannot measure the shape 
parameter $\alpha$ very well either and it also tends to be $1$. The mean and 
rms error are respectively $0.67$ and $0.23$, and the fractional error is 
$34.2\%$. Fig.~\ref{fig:agauss} shows the posterior probability density of the 
strain amplitude $\sigma_{\mathrm{GW}}$ and the spectrum power index 
$\alpha_{\mathrm{GW}}$ the gravitational wave background obtained by assuming 
the background is Gaussian and applying the conventional method in 
\cite{haasteren:2009:omt}.

\begin{figure}[ht]
\centering
\includegraphics[width=15cm]{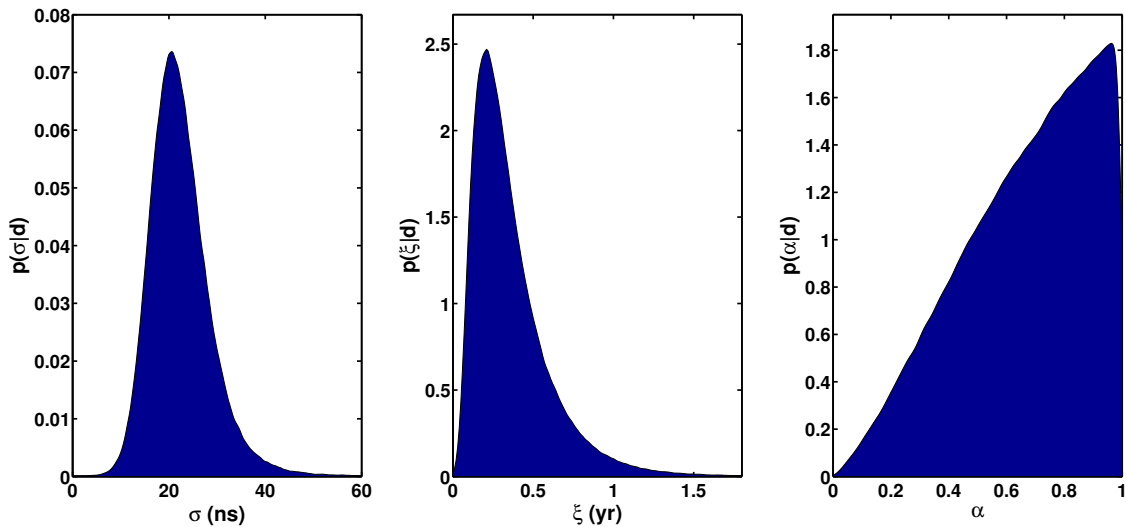}
\caption[Posterior of hyperparamters of the nonparametric model for the Gaussian 
gravitational wave background]{Posterior probability densities of 3 
hyperparameters --- $\sigma$, $\xi$ and $\alpha$, for analysis on the data 
described in Sec.~\ref{sec:anisotropic}. The fractional errors of them are 
respectively $30.0\%$, $71.9\%$, $34.2\%$}
\label{fig:ahyper}
\end{figure}

\begin{figure}[ht]
\centering
\includegraphics[width=15cm]{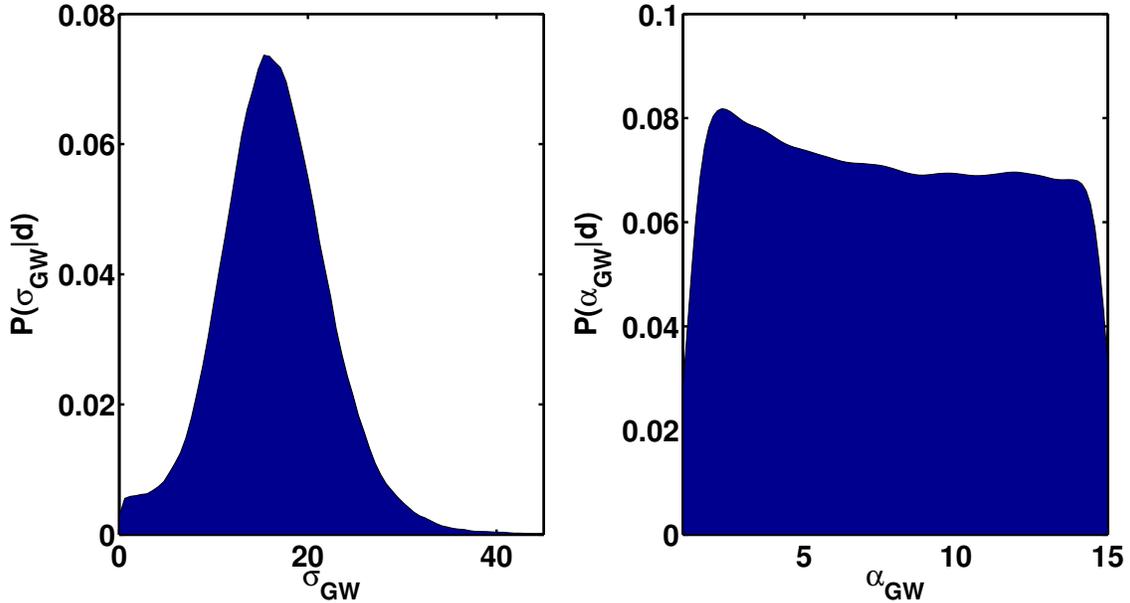}
\caption[Posterior of parameters of Gaussian parametric model for the Gaussian 
gravitational wave background]{Posterior probability densities the strain 
amplitude $\sigma_{\mathrm{GW}}$ and the spectrum power index 
$\alpha_{\mathrm{GW}}$ the gravitational wave background with $10^5$ sources 
obtained by assuming the background is Gaussian and applying the conventional 
method in \cite{haasteren:2009:omt}.}
\label{fig:agauss}
\end{figure}
\end{itemize}

\subsubsection{No Signal} \label{sec:none}
\begin{figure}[ht]
\centering
\includegraphics[width=15cm]{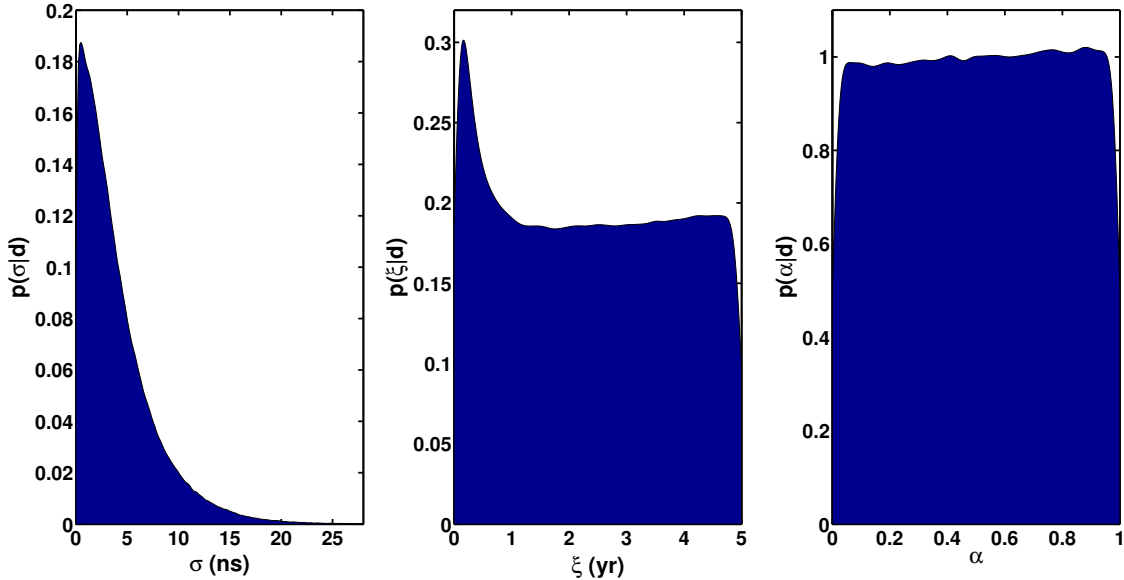}
\caption[Posterior of hyperparameters of nonparametric model for noise alone 
data]{Posterior probability densities of 3 hyperparameters --- $\sigma$, $\xi$ 
and $\alpha$, for analysis on the noise alone data described in 
Sec.~\ref{sec:none}. The fractional errors of them are respectively $92.5\%$, 
$63.5\%$, $54.6\%$}
\label{fig:nhyper}
\end{figure}

For comparative study, we also apply our Bayesian nonparametric analysis to a 
data set with timing noise alone. The third row of Table.~\ref{tab:results} and 
Fig.~\ref{fig:nhyper} summarize the results. The difference between the DICs of 
the two repulsive hypothesis is $5$, which shows that data favors the null 
hypothesis. All the hyperparameters are imprecisely determined. The 
method in \cite{haasteren:2009:omt} also offers a DIC difference of $5$.

% flatex input end: [demo.tex]

%\usepackage{hyperref}

% flatex input: [concl.tex]
\section{Conclusion} \label{sec:concl}
First detection of gravitational waves will open a new window of our universe 
complementary with the conventional electromagnetic astronomy. Due to their 
unique nature, to detect gravitational waves do not only requires more sensitive 
and innovative instruments, but it also demands more advanced analysis 
methodology and techniques. In this paper, we use a Bayesian nonparametric 
method to analyze the pulsar timing array data set which may contain 
contribution from a gravitational wave background originated from the 
superposition of gravitational waves generated by supermassive black hole 
binaries in the universe. .

When the number of the gravitational wave sources that significantly contribute to 
pulsar timing signals is small, the 
previous methods based on the assumption that the background spectrum is a power law
may be restrictive. In order to detect a generic gravitational wave background, we 
treat it as a deterministic process rather than a stochastic one as before, 
since each gravitational wave from a single source is a deterministic process. 
Instead of parameterizing gravitational wave from each single source, we use a 
different method --- Bayesian nonparametrics --- to avoid the 
over-parameterization. In this way, we set strong constraints on the feasible 
shapes of the pulsar timing residuals induced by the background. We have found 
that our method works efficiently for theoretically expected signals. When the number of 
gravitational wave sources becomes small and the assumption of power law spectrum becomes ineffective, 
our method is still able to detect and characterize the signal while the conventional method becomes less effective.  

% and it is not sensitive to the anisotropy of the background. 

For the purpose of demonstration, we apply our Bayesian nonparametric analysis 
to the pulsar timing data of the 4 best millisecond pulsars in current 
International pulsar timing array (IPTA), as the capability of detection and 
characterization of gravitational waves will be dominated by these pulsars 
\cite{burt:2011:opt}. However, our analysis can be straightforwardly applied to 
analyze the data of all the pulsars in IPTA. In the future, the effective number 
of pulsars whose timing noises are low enough to detect gravitational waves is 
expected to significantly increase with the birth of more sensitive radio 
telescopes such as Five-hundred-meter Aperture Spherical Telescope 
\cite{nan:2011:tfa} and Square Kilometer Array (SKA) \cite{dewdney:2009:tsk}. 
Applying our analysis method to the pulsar timing data collected by these future 
telescopes will significantly improve the detection sensitivity and inference of 
the signals. While the context of our discussion focuses on pulsar timing arrays, the 
analysis itself is directly applicable to detect and characterize any signals 
that arise from the superposition of a large number of astrophysical events, 
such as detecting high frequency gravitational wave background by LIGO 
\cite{abadie:2012:ulo}.

% flatex input end: [concl.tex]

%\usepackage{hyperref}

\begin{acknowledgments}
I thank my advisor Prof.~Lee Samuel Finn for fruitful discussions on Bayesian 
nonparametric analysis and valuable suggestions on the manuscript. This work was 
supported by Research Assistantship in the department of physics, and National 
Science Foundation Grant Numbers 09-40924 and 09-69857 awarded to The 
Pennsylvania State University.
\end{acknowledgments}

 \end{document}